\pdfoutput=1

\documentclass[aps,prl,twocolumn,superscriptaddress,a4paper,floatfix]{revtex4}
\usepackage[latin1]{inputenc}
\usepackage{graphicx}
\usepackage{amsmath,amssymb}
\usepackage{hyperref}
\usepackage{datetime}
\usepackage{upgreek}
\usepackage{placeins}

\usepackage{xcolor}
\renewcommand{\figurename}{Figure}
\renewcommand{\thetable}{\arabic{table}}

\makeatletter
\newcommand*{\newbibstartnumber}[1]{%
  \apptocmd{\thebibliography}{%
    \global\c@NAT@ctr #1\relax
    \addtocounter{NAT@ctr}{-1}%
  }{}{}%
}
\makeatother

\begin{document}

\title{Mapping out spin and particle conductances in a quantum point contact}

 \author{Sebastian Krinner}
 \affiliation{Department of Physics, ETH Zurich, 8093 Zurich, Switzerland}
 \author{Martin Lebrat}
 \affiliation{Department of Physics, ETH Zurich, 8093 Zurich, Switzerland}
 \author{Dominik Husmann}
 \affiliation{Department of Physics, ETH Zurich, 8093 Zurich, Switzerland}
 \author{Charles Grenier}
 \affiliation{Department of Physics, ETH Zurich, 8093 Zurich, Switzerland}
 \author{Jean-Philippe Brantut}
%  \email{brantutj@phys.ethz.ch}
 \affiliation{Department of Physics, ETH Zurich, 8093 Zurich, Switzerland}
 \author{Tilman Esslinger}
 \affiliation{Department of Physics, ETH Zurich, 8093 Zurich, Switzerland}

%\contributor{Submitted to Proceedings of the National Academy of Sciences
%of the United States of America}
%
%\significancetext{Predicting the transport properties of interacting particles in the quantum regime is challenging. A conceptually simple situation is realized by connecting two reservoirs through a quantum point contact. For non-interacting fermions, the conductance is quantized in units of the inverse of Planck's constant, reflecting the contribution to transport of an individual quantum state. We use a cold atomic Fermi gas to map out the spin and particle conductance in a quantum point contact for increasing attractive interactions, and observe remarkable effects of interactions on both spin and particle transport. Our work provides maps of conductance over a wide range of parameters and in the presence of many-body correlations, yielding new insights into the nature of strongly attractive Fermi gases.}

%\begin{article}
\begin{abstract}
We study particle and spin transport in a single mode quantum point contact using a charge neutral, quantum degenerate Fermi gas with tunable, attractive interactions. This yields the spin and particle conductance of the point contact as a function of chemical potential or confinement. The measurements cover a regime from weak attraction, where quantized conductance is observed, to the resonantly interacting superfluid. Spin conductance exhibits a broad maximum when varying the chemical potential at moderate interactions, which signals the emergence of Cooper pairing. In contrast, the particle conductance is unexpectedly enhanced even before the gas is expected to turn into a superfluid, continuously rising from the plateau at $1/h$ for weak interactions to plateaux-like features at non-universal values as high as $4/h$ for intermediate interactions. For strong interactions, the particle conductance plateaux disappear and the spin conductance gets suppressed, confirming the spin-insulating character of a superfluid. Our observations document the breakdown of universal conductance quantization as many-body correlations appear. The observed anomalous quantization challenges a Fermi liquid description of the normal phase, shedding new light on the nature of the strongly attractive Fermi gases. 
\end{abstract}

\maketitle
Quantum gas experiments provide a tool to study fundamental concepts in physics, which may be hard to access by other means. Challenges such as the interplay and dynamics of many interacting fermions, are addressed by interrogating a specifically tailored quantum many-body system with controlled parameters, an approach referred to as quantum simulation \cite{Cirac:2012aa}. The outcomes can then be used  to benchmark theory, or even as a direct comparison to different experimental realizations of the same concept. During the last years there has been substantial progress on this path using cold atomic gases to realize important models of condensed matter physics, formulated to describe the bulk properties of materials \cite{Esslinger:2010aa,Bloch:2012aa}. Here, neutral fermionic atoms are used to model the electrons in a solid. 

In this article we use a quantum gas to study the operation of an entire mesoscopic device, a quantum point contact (QPC), in the presence of interactions between the particles. We observe the transport of a charge neutral quantum degenerate gas of fermionic lithium atoms, which can be prepared in a mixture of two hyperfine states. These states provide a spin degree of freedom and the attractive interaction between them can be tuned continuously from weak to unitary, a feature unique to cold atomic gases. The QPC itself is realized by a suitably shaped optical potential, which consists of a short, one-dimensional channel connected to two large reservoirs \cite{van_wees_quantized_1988,wharam_one-dimensional_1988}. Biasing the reservoirs with different chemical potentials can induce a DC current. The ratio of the current to the bias is the conductance of the contact, which is independent of the bias in the linear response regime. 

For spinless, non-interacting particles at low temperature, the conductance is quantized in units of $1/h$, the universal conductance quantum for neutral particles \cite{krinner_observation_2015}. An intuitive understanding thereof can be gained by considering the temporal spacing $\tau = h/\Delta \mu$ of minimum uncertainty wave packets within one transverse mode, and moving through the channel in response to an applied bias $\Delta \mu$ \cite{PhysRevB.45.1742,Batra:1998aa,PhysRevB.78.165330,PhysRevLett.108.186806}. 
At zero temperature, within one energetically available mode, each wavepacket state is occupied by a single particle in accordance to Pauli's principle. Hence the maximum current carried by the mode is $I=1/\tau=\Delta\mu/h$ with $1/h$ being the upper bound for the contribution of a single mode to conductance. This bound, set by Heisenberg's and Pauli's principle, holds for all systems where transport proceeds by fermionic quasiparticles, such as Fermi liquids.

In a gas with two spin components, the conductance of each component is quantized in units of $1/h$, as long as there are no interactions between them. The picture becomes more intricate if collisions between the two components play a role. Studies conducted on solid-state systems have shown that the weak correlations in conventional superconductors yields quantized supercurrents \cite{Beenakker:1991aa,PhysRevB.71.174502} and the emergence of Andreev bound states in mesoscopic conductors\cite{Kleinsasser:1994aa,PhysRevLett.85.170,Bretheau:2013aa}. Besides, in semiconductor systems, the Coulomb repulsion between charge carriers subtly modifies the conductance quantization \cite{PhysRevLett.77.135,Cronenwett:2002fk,PhysRevLett.92.106801,bauer_microscopic_2013,Iqbal:2013aa}, lowering conductance at low density and leading to strong correlations. 

To characterize the transport, one may additionally consider the spin current, i.e. the relative current of one component with respect to the other, which is expected to be damped in the presence of interactions \cite{mink_spin_2012}. While measuring and inducing spin currents in clean mesoscopic conductors is a challenge in solid state systems, quantum gases naturally allows for spin resolved observations and manipulations. For example the damping of spin currents was measured in strongly interacting Fermi gases at high temperature, where many-body effects, in particular pairing, are weak \cite{sommer_universal_2011, koschorreck_universal_2013, bardon_transverse_2014}. In contrast, the total particle current is conserved in collisions, thus the quantization of particle conductance in the channel should be robust. In fact, it was shown, that the applicability of a Fermi liquid description in the leads of the contacts guarantees universal conductance quantization \cite{Safi:1995aa,Ponomarenko:1995aa,maslov_landauer_1995}, regardless of the interaction strength. 

Our study of spin and particle conductance of a quantum point contact with tunable interactions uses an atomic Fermi gas in the vicinity of a broad Feshbach resonance, realising the BEC-BCS crossover regime \cite{Zwerger:2011aa}. It features a conventional s-wave paired superfluid for strong attraction and low temperatures, as well as a Fermi-liquid phase for weak attraction and high temperatures. The nature of the state in the intermediate regime remains controversial, as it is governed by a non-trivial interplay of pairing and superfluid fluctuations competing with finite temperature properties of the gas \cite{sagi_breakdown_2015}. This richness makes our system an ideal test bed to study how transport properties change with interactions. On the one hand, a spin-insulating character, not accessible to high temperature measurements \cite{sommer_universal_2011}, should emerge as a result of s-wave pairing. On the other hand, particle transport directly tests the robustness of conductance quantization as many-body correlations emerge. 

\section{System}

\begin{figure*}
\centerline{\includegraphics[width=178mm]{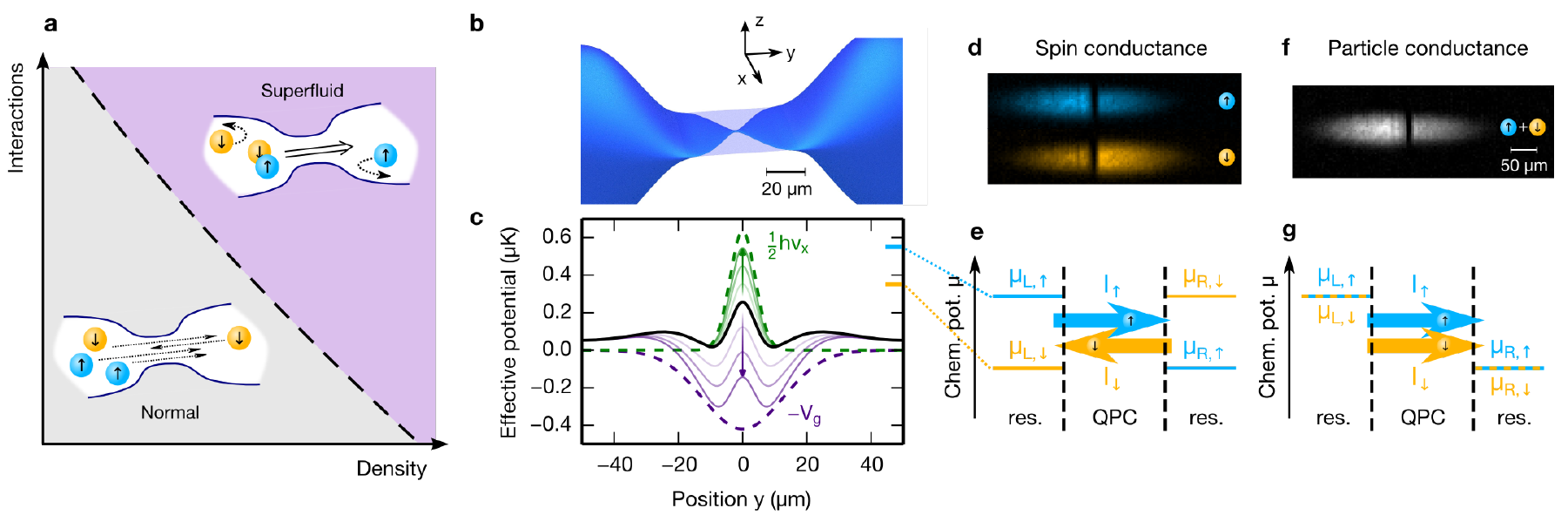}}
    \caption{{\bf Concept of the experiment}
    {\bf a,} Low-temperature phase diagram of the attractive Fermi gas at fixed temperature. In the normal, weakly-interacting phase the two spin components move independently of each other in the QPC. In the superfluid phase large particle currents arise, whereas spin currents are strongly suppressed due to pairing.
    {\bf b,} Three-dimensional impression of the QPC, connected via an intermediate 2D region to large 3D reservoirs (only shown partly).
    {\bf c,} Effective potentials in the central region around the QPC along the transport axis $y$. It is the sum of the zero-point energy of the QPC (green dashed line: contribution from confinement along $x$), an attractive gate potential (purple dashed line) and the underlying harmonic trap (see Materials and Methods). The black solid line corresponds to the parameters for which the conductance plateau in Fig. \ref{fig2}a is observed. Thin violet lines show how the effective potential evolves when $V_g$ is increased from $0.42\,\mu$K to $0.82\,\mu$K, whereas thin green lines depict the corresponding evolution when $\nu_x$ is increased from $13.2\,$kHz to $25.2\,$kHz. Note the two local minima of the effective potential at the entrance and exit of the QPC.
    {\bf d,} Absorption images of the $\uparrow$ and $\downarrow$ cloud components as prepared before spin conductance measurements.
    {\bf e,} Chemical potentials and currents in the presence of a spin bias.
    {\bf f,} Absorption image of the atoms prepared for the particle transport, with identical bias for $\uparrow$ and $\downarrow$.
    {\bf g,} Chemical potentials and currents in the presence of a chemical potential bias.
    }
    \label{fig1}
\end{figure*}

\subsection{Implementation}

We capture in an elongated harmonic trap a total of $N=9.6(3)\times10^4$ $^6\textup{Li}$ atoms in each of the lowest and third lowest hyperfine states, labeled $\downarrow$ and $\uparrow$. The particles interact via the van der Waals potential, which at the relevant density and energy scales reduces to a contact interaction characterised by the s-wave scattering length $a$. The scattering length, controlling the interaction strength, is adjusted by setting a homogeneous magnetic field between 673 and 949\,Gauss, covering the regime from $1/k_{\textup{F,res}}a=-2.0$ to 0.6, where $k_{\textup{F,res}}=\sqrt{2 m E_F/\hbar^2}$ is the Fermi wavevector in the gas, $m$ the mass of $^6\textup{Li}$ atoms and $E_F=k_B T_F=\hbar\bar{\omega}(6N)^{1/3}$ the Fermi energy in the harmonic trap, with $\bar{\omega}$ being the geometric mean of its frequencies. We reach temperatures of 0.15(2)\,$T_F$ for the strongest and 0.11(2)\,$T_F$ for the weakest interactions (SI Text). For strong interaction, the temperature is low enough to access the superfluid regime, as sketched in figure \ref{fig1}a.
Starting from the trapped gas, we first imprint a two-dimensional constriction at the center of the cloud using an off-resonant laser beam operating at a wavelength of 532\,nm and shaped in a $\textup{TEM}_{01}$-like mode propagating along the $x$ axis and hitting the cloud at its centre. This separates the elongated cloud into two reservoirs smoothly connected by a quasi-two dimensional region, with a maximum vertical trap frequency along $z$ of $\nu_z=9.2(4)$\,kHz. The quantum point contact itself, depicted in Fig.~\ref{fig1}b, is created by imaging a split gate structure on the two-dimensional (2D) region using high-resolution lithography \cite{krinner_observation_2015}. It is characterized by its transverse trapping frequency at the centre, $\nu_x$, which is adjustable. The Gaussian envelopes of the beams ensure a smooth connection to the large three-dimensional reservoirs formed at both ends of the cloud, see Fig.~\ref{fig1}b.
An attractive gate potential $V_g$ is realized by a red-detuned laser beam focused on the QPC \cite{krinner_observation_2015} which tunes the chemical potential in the QPC and its immediate vicinity, see Fig.~\ref{fig1}c. The left (L) and right (R) reservoirs connected to the QPC contain $N_{i,\sigma}$ atoms and have chemical potentials $\mu_{i,\sigma}$, with $i$ = $L$, $R$ and $\sigma$ = $\uparrow$, $\downarrow$. Because the gate potential has a waist larger than the QPC, it also increases the density at the entrance and exit points (minima of the effective potential in Fig.~\ref{fig1}c), creating a dimple effect that increases the local degeneracy \cite{Stamper-Kurn:1998aa}.
\subsection{Initialization}
To measure the particle or spin conductances of the QPC, we prepare either an atom number imbalance $\Delta N = (\Delta N_{\uparrow} + \Delta N_{\downarrow})/2 \simeq 0.4\,N$ or a magnetization imbalance $\Delta M = (\Delta N_{\uparrow} - \Delta N_{\downarrow})/2 \simeq 0.25\,N$, with $\Delta N_{\sigma}= N_{L,\sigma} - N_{R,\sigma}$. These correspond to a chemical potential bias $\Delta\mu = (\Delta\mu_\uparrow+\Delta\mu_\downarrow)/2\simeq0.21(2)\mu\ll h\nu_z$ or a spin bias $\Delta b = (\Delta\mu_\uparrow-\Delta\mu_\downarrow)/2\simeq0.24\mu$ respectively, see Fig.~\ref{fig1}d-g and Materials and Methods. In the weakly interacting regime, we do not find deviations from linear response within our experimental uncertainties \cite{krinner_observation_2015}. For particle transport in the strongly interacting regime, $|1/k_{\textup{F,res}} a| < 0.7$, we observe the emergence of non-linearities allowing us to identify the superfluid regime (see \cite{Husmann:2015aa} and SI Text). 
The interaction-dependent chemical potentials $\mu_{i,\sigma}$, and the chemical potential at equilibrium $\mu$ are extracted from the known equation of state of the tunable Fermi gas \cite{navon_equation_2010} (see Materials and Methods).
The biases induce a spin current $I_\sigma$ and a particle current $I_N$ defined as
\begin{equation}
\begin{cases}
I_{\sigma} = -\frac{1}{2}\frac{\textup{d}}{\textup{d}t}\Delta M = G_{\sigma} \Delta b \\
I_{N} = -\frac{1}{2}\frac{\textup{d}}{\textup{d}t}\Delta N  = G_N \Delta \mu 
\end{cases}
\label{eq:linearresp}
\end{equation}
where $G_\sigma$ and $G_N$ are the spin and particle conductances, respectively. The currents are estimated by measuring the number of particles transferred after $4$\,s and $2$\,s of transport time respectively, and compared to the bias to obtain the conductances (see Materials and Methods).

\begin{figure}[htdp]
\centerline{\includegraphics[width=80mm]{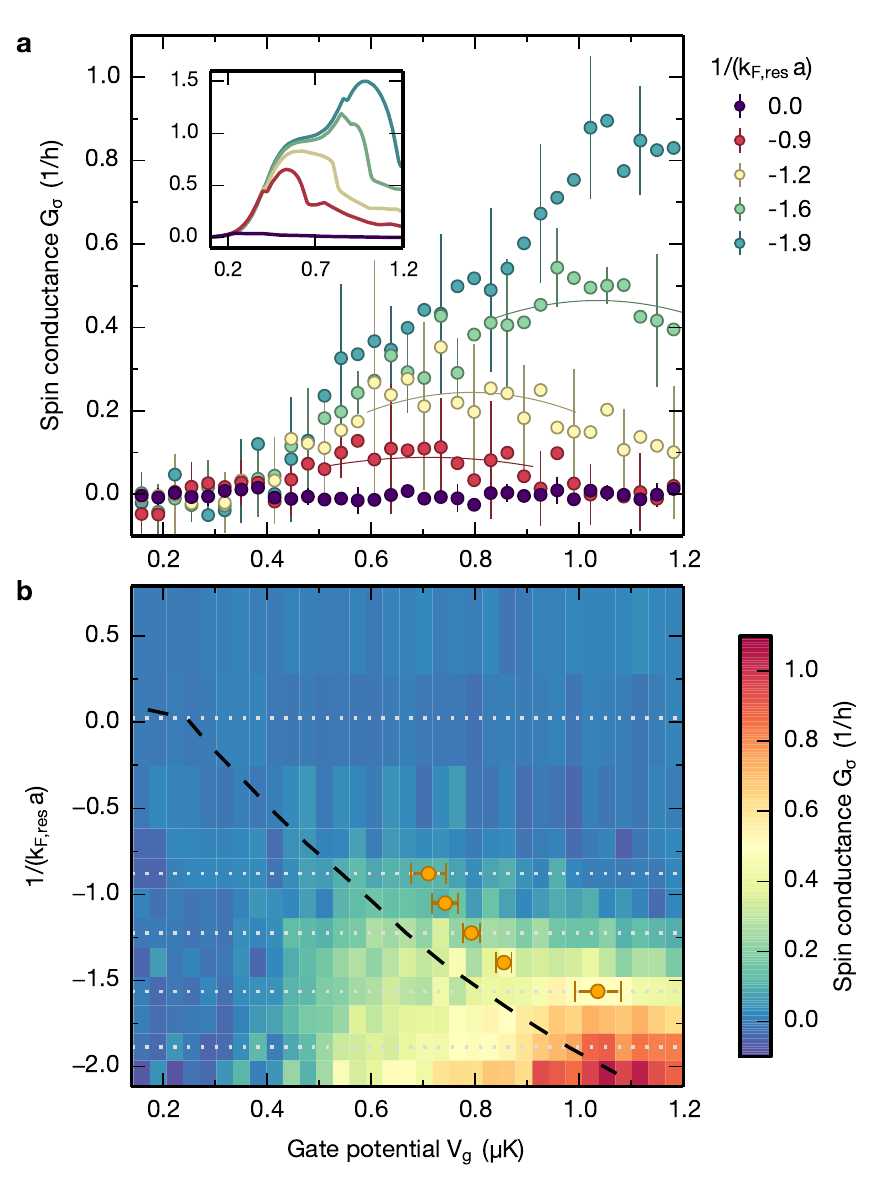}}
    \caption{{\bf Spin conductance of the attractively interacting Fermi gas.} {\bf a,} 
    Spin conductance $G_\sigma$ as a function of the gate potential $V_g$ for different interaction strengths $1/(k_{\textup{F,res}} a)$ in the reservoirs. Each data point represents the mean over 9 measurements and error bars indicate one standard deviation plotted for every third point. The thin solid lines are quadratic fits used to identify the maxima in $G_\sigma$. Inset: $G_\sigma$ obtained from a mean-field phenomenological model, reproducing the non-monotonic behaviour of the experimental data.
    {\bf b,} Two-dimensional color plot of $G_\sigma$ as a function of $1/(k_{\textup{F,res}} a)$, with cuts of Fig. 2a indicated as grey dotted lines.
    The points where $G_\sigma$ is maximum, obtained from a parabolic fit along $V_g$, are displayed as orange circles for comparison. The black dashed line represents the superfluid critical line estimated at the entrance and exit regions of the QPC, using the results of \cite{haussmann_thermodynamics_2007}.}
    \label{fig3}
\end{figure}

\section{Spin transport}

\subsection{Measurements}
We first investigate the spin conductance $G_{\sigma}$ as a function of gate potential $V_g$, at fixed $\nu_x=23.2(2.5)\,$kHz, but for different interaction strengths. For non-interacting particles, the strengths of the gate determines the number of transport channels that are open \cite{krinner_observation_2015}.
The results are presented in Fig.~\ref{fig3}a and b. For the weakest interactions, we observe the onset of spin transport as the first channel opens around $V_g=0.4\,\mu$K, followed by a continuous increase of $G_\sigma$ up to the largest gate potentials. The second channel is expected to open around $V_g= 0.8\,\mu$K.
For intermediate interaction strengths $-1.7 < (1/k_{\textup{F,res}} a) < -1.0$, we observe a broad maximum in $G_\sigma$ as a function of $V_g$.
The opening of the channel is still indicated by a sharp increase of $G_\sigma$ at an interaction-independent value of the gate potential. 
With increasing interactions, the centre of the broad maximum in $G_\sigma$ shifts to lower $V_g$, and its height is reduced. For $1/(k_{\textup{F,res}} a) > -0.5$, $G_\sigma$ vanishes over the entire range of gate potentials. A complete map of $G_\sigma$ as a function of interaction strength and gate potential is shown in Fig.~\ref{fig3}b.
The existence of a maximum in $G_\sigma$ and the negative spin transconductance $\textup{d}G_{\sigma}/\textup{d}V_g$ for strong interactions indicate the appearance of a spin insulating phase. Indeed, increasing $V_g$ increases the chemical potential in and around the QPC, reducing the relative temperature $T/T_F$. The decrease of conductance with decreasing temperature is characteristic of an insulating behaviour \cite{Gebhard:2003aa}.

\subsection{Mean-field model}

We use a mean-field approach to capture the phenomenology of the spin transport. It assumes that excitations are non-interacting, fermionic Bogoliubov quasi-particles. Since the Cooper pairs are singlets, these excitations carry the spin current and their populations are controlled by the spin bias. This allows for a generalisation of the Landauer approach to spin conductance (SI text). The predictions are shown in the inset of Fig~\ref{fig3}a. The emergence of a maximum as a function of gate potential is reproduced. It results from the competition between the non-linearly increasing gap at the entrance and exit of the QPC, hindering spin transport, and the opening of conduction channels.

The position of the maximum along the gate potential axis, and the shape of the conductance variations are reproduced, but the predictions for the value of the conductance differ by about a factor of two. This discrepancy could come from interactions between quasi-particles of opposite spin, neglected in the model, in particular inside the contact where the one dimensional geometry enhances scattering \cite{olshanii_atomic_1998}.

\section{Particle transport}
\begin{figure*}
    \centerline{\includegraphics[width=114mm]{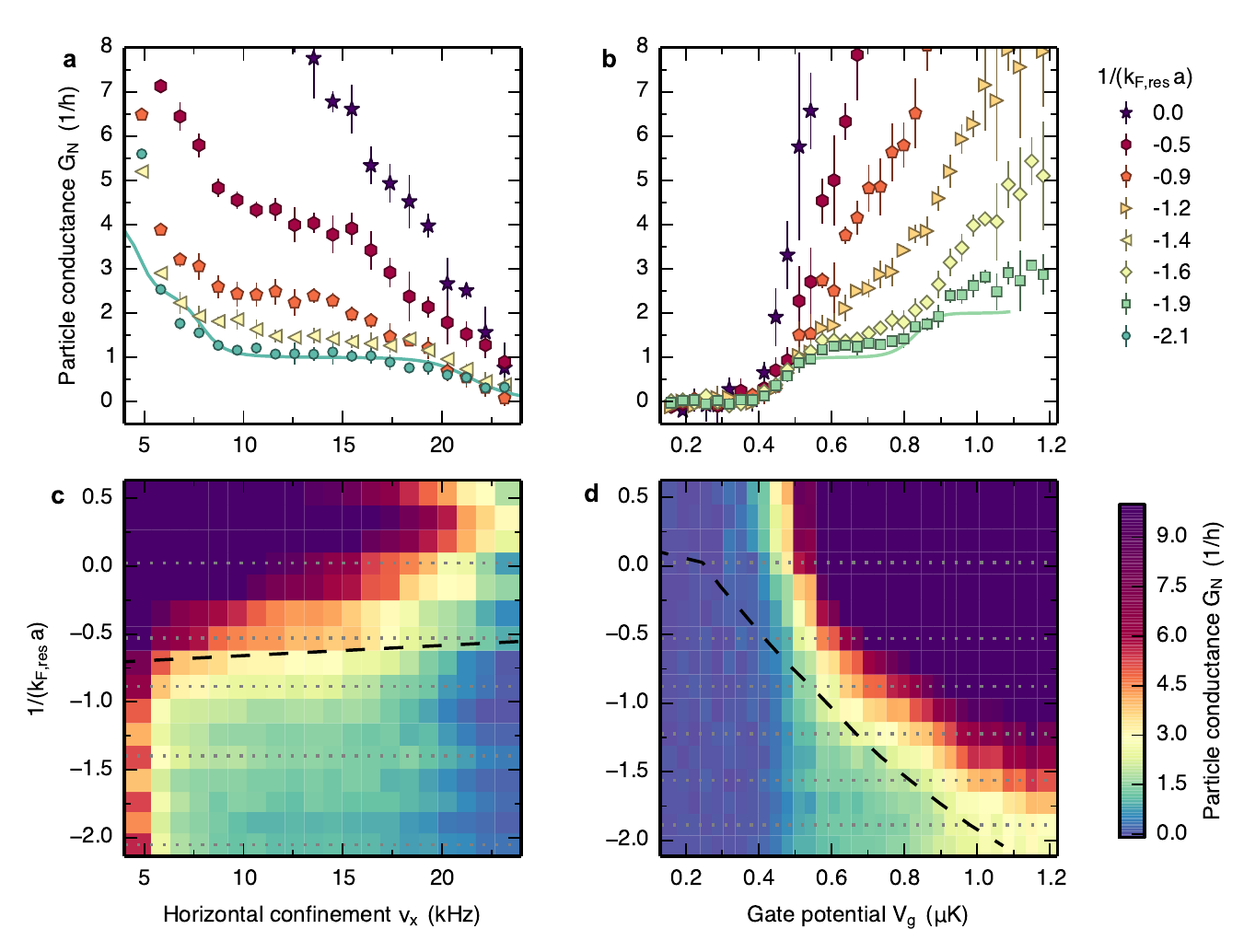}}
    \caption{{\bf Particle conductance of the attractively interacting Fermi gas.} {\bf a,} Particle conductance $G_N$ as a function of the horizontal confinement frequency $\nu_x$ of the QPC, at fixed gate potential $V_g = 0.42\,\mu$K; and {\bf b,} as a function of the gate potential $V_g$ at fixed confinement frequency $\nu_x = 23.2$kHz, for different interaction strengths $1/(k_{\textup{F,res}} a)$ in the reservoirs. The solid lines are theoretical predictions for $1/(k_{\textup{F,res}} a) = 2.1$ and 1.9 respectively, based on the Landauer formula including mean-field attraction (SI Text). Each data point represents the mean over 5 measurements and error bars indicate one standard deviation. {\bf c,} and {\bf d,} Two-dimensional colour plot of $G_N$ as a function of interaction strength $1/(k_{\textup{F,res}} a)$ and horizontal confinement (c) or gate potential (d). Both plots contain the cuts of Fig. 3a and b (grey dotted lines), and an estimation of the local superfluid transition at the QPC exits (black dashed line).
    }
    \label{fig2}
\end{figure*}

\subsection{Measurements}

The total particle current is expected to be robust against collisions, since they conserve momentum. Yet, large interaction effects are observed. 
We measured $G_N$ for several interaction strengths as functions of $V_g$ and $\nu_x$. Fig. \ref{fig2}a shows the curves for fixed gate potential $V_g=0.42\,\mu$K as a function of horizontal confinement $\nu_x$.
For the weakest interaction strength $1/k_{\textup{F,res}}a=-2.1$, $G_N$ shows a distinct plateau at $1/h$ in agreement with the Landauer picture. For the tightest horizontal confinement, $\nu_x = 23.2\,$kHz, the QPC is almost pinched off, while when reducing $\nu_x$ below 8\,kHz, several transverse modes with closely spaced energies get populated.

For interaction strengths $-2.1 < 1/(k_{\textup{F,res}}) a < -0.5$, a conductance plateau with a reduced length remains visible in Fig. \ref{fig2}a. The height of this feature continuously increases above the universal value, and eventually washes out with increasing interaction strength, leaving a visible shoulder as high as $\sim4/h$ for $1/(k_{\textup{F,res}} a) = -0.5$.
A similar observation is made when varying $V_g$ at fixed $\nu_x=23.2$\,kHz, as shown in Fig. \ref{fig2}b. There again, plateaux-like features with conductances higher than $1/h$ are observed for interaction strengths $1/(k_{\textup{F,res}} a) < -1.3$. As interactions are further increased towards the unitary regime ($-0.5 < 1/(k_{\textup{F,res}} a) \leq 0$ for Fig. \ref{fig2}a and c, $-1.3 < 1/(k_{\textup{F,res}} a) \leq 0$ for Fig. \ref{fig2}b and d), no conductance plateaux can be distinguished, and $G_N$ increases continuously from zero to large values. Contrary to variations of $\nu_x$, variations of $V_g$ change the density at the entrance and exit of the QPC, which probably causes the disappearance of the plateau already at a lower value of the interaction strength. 

The entire crossover from quantized conductance of weakly interacting atoms to its breakdown for strong interactions is mapped out in Fig. \ref{fig2}c for fixed $V_g$ and varying $\nu_x$, and in Fig. \ref{fig2}d for fixed $\nu_x$ and varying $V_g$. The latter demonstrates most clearly that the conductance plateau, discernible as green area, shrinks gradually when the interaction strength is increased from $1/(k_{\textup{F,res}} a) < -2$ to $1/(k_{\textup{F,res}} a) < -1$. In this regime the plateau width is well predicted by a mean-field model accounting for intra- and inter-mode attraction in the QPC (SI Text). Furthermore, we observe little difference between the unitary and the molecular regime in the experimentally accessible region, $0 < 1/(k_{\textup{F,res}} a) < 0.5$, where the reservoirs form a condensate of molecules \cite{Giorgini_RMP_2008}.

\subsection{Superfluid transition}
In the strongly interacting regime (deep purple regions in Fig. \ref{fig2}c and d), deviations from a linear response to the bias are observed \cite{Jendrzejewski:2014aa,Labouvie:2015aa} in agreement with our previous measurements for a QPC in a unitary superfluid \cite{Husmann:2015aa} (SI Text). Indeed, for the temperature imposed by the reservoirs, increasing $V_g$ or the interactions leads to the onset of superfluidity in the minima of the effective potential (see Fig. \ref{fig1}c), i.e. at the entrance and exit of the QPC. The local critical temperature at those points thus corresponds to the maximum critical temperature over the entire cloud, and we refer to it as $T_c$ for the remainder. To extract it, we use the state-of-the-art calculation of $T_c/\tilde{T}_{\textup{F}}\left(1/(\tilde{k}_{\textup{F}} a)\right)$ \cite{haussmann_thermodynamics_2007} in local density approximation, with $k_B \tilde{T}_{\textup{F}}=\hbar^2\tilde{k}_{\textup{F}}^2/(2m)=\hbar^2(6\pi^2 n)^{2/3}/(2m)$ being the Fermi energy of a homogeneous gas with density $n$. We estimate $n$ at the entrance and exit of the QPC from the trap geometry and the equation of state of the low-temperature, tunable Fermi gas (SI Text). The resulting critical line is displayed in Fig. \ref{fig3}b, \ref{fig2}c and d. It closely tracks the maxima of the spin conductance in Fig. \ref{fig3}b, as well as the disappearance of the conductance plateaux in Fig. \ref{fig2}. 

\subsection{Conductances in the single mode regime}

\begin{figure}%[htdp]
    \centerline{\includegraphics[width=80mm]{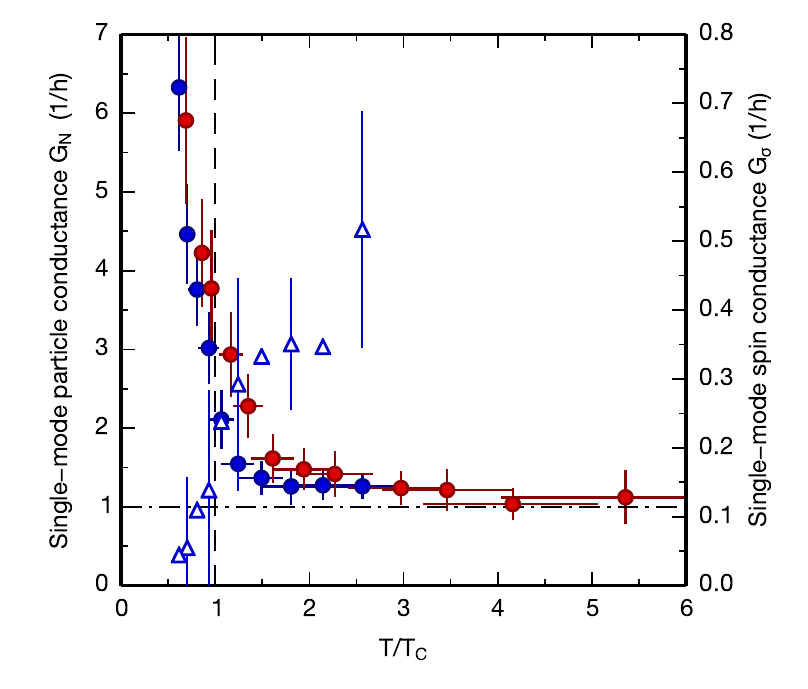}}
    \caption{{\bf Particle and spin conductances in the single mode regime}.
    $G_N$ (closed circles) and $G_\sigma$ (open triangles, every second error bar displayed) for various interaction strengths are presented as a function of the reduced temperature $T/T_c$, which varies due to the dependence of $T_c$ on density and scattering length.
    Blue data points are obtained from the measurements shown in Fig. 2 and Figs. 3b, d, for $V_g=0.64\mu$K and $\nu_x=23.2$\,kHz.
    Red data points are obtained from the measurements shown in Figs. 3a, c, for $V_g=0.42\mu$K and $\nu_x=14.5$\,kHz.
    $G_N$ tends to the conductance quantum $1/h$ (horizontal dash-dotted line) for weak interactions ($T/T_c\gg 1$). Error bars contain statistical and systematic errors (see Materials and Methods).}
    \label{fig5}
\end{figure}

We now focus on the conductances in the single mode regime, where universal quantization is observed for weak interactions. For this purpose, we display in Fig. \ref{fig5} the conductances as a function of $T/T_c$, measured at the position of the plateau center in the weakly interacting regime. These are extracted from Fig. \ref{fig2}c for fixed $\nu_x=14.5$\,kHz and from Fig. \ref{fig2}d for fixed $V_g=0.64\mu$K. We observe that the resulting conductances now coincide within error bars. This demonstrates that $T/T_c$ is a key control parameter of the transition, despite the fact that the two data sets correspond to different geometries in the single mode regime. The fast increase of particle conductance coincides with a sharp drop in the spin conductance around $T/T_c=1$, demonstrating directly the intimate connection between pairing and superfluidity. The regime of non-universal quantisation, with a conductance larger than $1/h$ as identified by our measurement method and accuracy, extends from $T/Tc\sim1$, corresponding to the shoulder observed at $\sim 4/h$, to far above the superfluid transition, up to $T/T_c \sim 2.5$. This suggests that in this regime, where $T>T_c$ at every point in the cloud, current is not carried by fermionic quasiparticles, challenging a description in terms of Fermi liquids. 

\section{Discussion}
A possible interpretation for the anomalously high conductance, i.e. exceeding $1/h$ in the normal, single mode regime, is the presence of strong superfluid fluctuations in the reservoirs, due to the large critical region around the superfluid transition \cite{Taylor:2009aa,Debelhoir:2015aa}. The critical fluctuations are qualitatively similar to that of the Luttinger liquid in one dimension with attractive interactions in the leads, where they are known to yield an enhanced conductance \cite{maslov_landauer_1995,Ponomarenko:1995aa,Safi:1995aa}. Another possibility are preformed pairs above $T_c$, that could form in particular in the contact region \cite{Moritz:2005aa}, leading to a channel with a bosonic character, where large conductances are expected \cite{Papoular:2015aa,Lee:2015aa}. Evidence for such non-Fermi liquid behaviour in the BEC-BCS crossover was found using photoemission spectroscopy \cite{sagi_breakdown_2015}, in contrast to results based on the equation of state \cite{Nascimbene:2011aa,ku_revealing_2012}.
Our findings, covering the attractively interacting regime, complement the observations made with repulsively interacting electrons in solid state QPCs. Future work could also explore the known conductance anomalies observed in electronic QPCs \cite{bauer_microscopic_2013,Iqbal:2013aa}.

\section{Materials}
\section{Preparation of the cloud}
Interacting Fermi gases are produced by a two steps evaporative cooling procedure. We first create a balanced mixture of the lowest two hyperfine states of $^6\textup{Li}$, and perform evaporative cooling at a magnetic field of $302$\,G down to temperatures of the order of the Fermi temperature. A Landau-Zener radio frequency transition then transfers the full population from the second to the third hyperfine state, and the magnetic field is ramped up to $689$\,G, the centre of a Feshbach resonance \cite{zurn_precise_2013}, where the s-wave scattering length diverges. A second step of forced evaporation is then performed using a magnetic field gradient, yielding the low temperature clouds used for the transport measurements. The magnetic field is then ramped in $200\,$ms to the desired value between 673 and 949\,Gauss in order to vary the interaction strength during transport. In the absence of the point contact and gate beam, the atoms reside in a hybrid trap where the confinement along $x$ and $z$ is ensured by an optical dipole trap, and along $y$ by the residual curvature of the magnetic field. The trap frequencies along the $x$ and $z$ directions are $194\,$Hz and $157\,$Hz respectively. The trap frequency along $y$ ranges between $28.0(5)\,$Hz and $33.2(6)\,$Hz, depending on the value of the magnetic field. 

\section{Creation of the spin bias}
To create a symmetric spin bias $\Delta\mu_{\uparrow}=-\Delta\mu_{\downarrow}$ between the two 
reservoirs (see Fig.~\ref{fig1}e), we ramp the magnetic field before the Landau-Zener transfer in 10\,ms 
from 302\,G down to 52\,G, where the lowest and second lowest hyperfine states have different magnetic 
moments, and apply at the same time a magnetic field gradient along the transport axis. This induces 
dipole oscillations with different frequencies and different amplitudes for the two states. We wait 
for roughly one period of the faster oscillation before we abruptly switch on an elliptic repulsive 
gate laser beam separating the reservoirs \cite{brantut_thermoelectric_2013}. The magnetic 
field is then ramped back to a value close to its initial value, from where we transfer all the atoms in state $|2\rangle$ to 
state $|3\rangle$ using an adiabatic Landau-Zener radio frequency transfer. After evaporation at the Feshbach 
resonance, we obtain an opposite atom number imbalance for the two states, $\Delta N_{\uparrow}\simeq-\Delta N_{\downarrow}\simeq 0.25$. 
We ensured that this preparation scheme does not increase the temperature as 
compared to the one for the particle transport.

\section{Effective potential}
For the computation of the conductance we use of adiabatic approximation \cite{glazman_reflectionless_1988, ihn_nanostructures}, which allows for a separation of longitudinal ($y$) and transverse ($x$, $z$) coordinates. %It neglects scattering between different transverse modes and is justified if the confinement of the constriction varies smoothly along the transport direction. 
We verified numerically that it is a very good approximation for the geometry of our QPC. In the resulting one-dimensional Schr\"{o}dinger equation the transverse energy $E_\perp(y)=\frac{1}{2}h\nu_xf_x(y)+\frac{1}{2}h\nu_zf_z(y)$ acts as an additional potential, with $f_{x,z}(y)$ describing the spatial variation of the trapping frequencies of the QPC. Together with the gate potential $V_g(y)=-V_g f_g(y)$, and the harmonic trapping potential $V_{\textup{trap}}(y)=\frac{1}{2}m\omega_y^2 y^2$ along $y$, and a residual repulsive potential $E_{\textup{resid}}(y)=E_{\textup{resid},0}f_z^2(y)$ arising from residual light in the nodal line of the intensity profile of the $\textup{TEM}_{01}$-like laser mode creating the 2D confinement \cite{brantut_conduction_2012}, it forms the effective potential $V_{\textup{eff}}= E_\perp + V_g+ V_{\textup{trap}} + E_{\textup{resid}}$,
which is drawn in Fig. \ref{fig1}c. The involved envelope functions are listed in Table \ref{envelopefunctions}. The prefactor in $E_{\textup{resid}}$ has been calibrated to $E_{\textup{resid},0}=0.14(7)\,\upmu$K using a conductance measurement with only the 2D confinement present. The central maximum in a generic profile $V_{\textup{eff}}$ is due to the $x$ confinement of the QPC, and the two minima to each side of it are a result of the combined potential of $E_\perp$ and $V_g$. We define the entrance and exit of the QPC as the position of these minima. They represent the positions of highest density and thus of lowest $T/\tilde{T}_{\textup{F}}$. 

\begin{table}[htdp]
\caption{Envelope functions determining the effective potential.}
\begin{tabular}{ccc}
Envelope function & Waist & Description\\
\hline
$f_x(y) = \textup{exp}(-y^2/w_x^2)$ & $w_x=5.6(6)\,\upmu$m & QPC, $x$ conf. \\
$f_z(y) = \textup{exp}(-y^2/w_z^2)$ & $w_z=30(1)\,\upmu$m & QPC, $z$ conf. \\
$f_g(y) = \textup{exp}(-2y^2/w_g^2)$ & $w_g=25(1)\,\upmu$m & Gate potential \\
\hline
\end{tabular}
\label{envelopefunctions}
\end{table}
\section{Compressibility and spin susceptibility of the trapped gas}
To evaluate the interaction dependent chemical potentials $\mu_\uparrow$ and $\mu_\downarrow$, compressibility $\kappa$ and spin bias $\Delta b$ we use the equation of state of the two component, homogeneous Fermi gas $\mathcal{P}(\mu_\uparrow,\mu_\downarrow,a)$ ~\cite{navon_equation_2010}. 
%It can be written as:
%\begin{equation}
% \mathcal{P}(\mu_\uparrow,\mu_\downarrow,a) = \mathcal{P}_0(\mu_\uparrow)\cdot h(\delta = \frac{\hbar}{\sqrt{2M\mu_\uparrow}a},\eta = \frac{\mu_\downarrow}{\mu_\uparrow})
%\end{equation}
%with $\mathcal{P}_0$ the equation of state of an ideal one component Fermi gas, $\mu_\uparrow$ and $\mu_\downarrow < \mu_\uparrow$ 
%the chemical potentials of the two species, and $h$ a function taking into account interactions and population imbalance.
Integrating it over the trap provides the thermodynamic potential $\mathcal{P}_\text{trap}= \int d\vec{r}\, \mathcal{P}(\mu_\uparrow - V(\vec{r}),\mu_\downarrow - V(\vec{r}),a)$,
%\begin{equation}
% \mathcal{P}_\text{trap} = \int d\vec{r}\, \mathcal{P}(\mu_\uparrow - V(\vec{r}),\mu_\downarrow - V(\vec{r}),a)\vartheta(\mu_\uparrow-V(\vec{r}))\,,
%\end{equation}
with $V(\vec{r})$ the known trapping potential (including the QPC region) and $\vartheta$ the Heaviside function. The particle number $N$ and the magnetization $M$ in a single reservoir are then given by $N = \frac{1}{2}\left(\frac{\partial \mathcal{P}_\text{trap}}{\partial \mu}\right)_b$ and $M = \frac{1}{2}\left(\frac{\partial \mathcal{P}_\text{trap}}{\partial b}\right)_\mu$
with $ \mu = \frac{\mu_\uparrow+\mu_\downarrow}{2}$ and $
b = \frac{\mu_\uparrow - \mu_\downarrow}{2}$. 
The factors of $1/2$ arise because the size of the two identical reservoirs is half of the entire cloud. 
Given the measured $N$ and $M$, one can solve numerically for $\mu_\uparrow$ and $\mu_\downarrow$, or equivalently, for $\mu$ and $b$.

The compressibility and spin susceptibility of a single reservoir are given by 
$\kappa = \frac{1}{4}\left(\frac{\partial^2 \mathcal{P}_\text{trap}}{\partial^2\mu}\right)_b$ and 
$\chi = \frac{1}{4}\left(\frac{\partial^2 \mathcal{P}_\text{trap}}{\partial^2 b}\right)_\mu$, respectively. The additional factors of 1/2 the definitions of $N$ and $M$ arise because in our definitions of $\kappa$ and $\chi$ we require that $\kappa\rightarrow(\partial N_{\uparrow(\downarrow)}/\partial\mu_{\uparrow(\downarrow)})_{b}$ for $b\rightarrow 0$.%, following the convention in cold atoms to state atom numbers and thermodynamic quantities of a 50:50 spin mixture only for a single spin component.

%The spin bias is defined as $\Delta b=(\Delta\mu_\uparrow-\Delta\mu_\downarrow)/2=\left[(\mu_{L,\uparrow}-\mu_{R,\uparrow})-(\mu_{L,\downarrow}-\mu_{R,\downarrow})\right]$. 
For a symmetric spin bias, %$\Delta\mu_\uparrow=-\Delta\mu_\downarrow$, as we apply it for the spin measurements, it is given by 
we have $\Delta b=2b$. It ranges from $0.18\mu$ to $0.34\mu$ for interaction strengths $-2.0 \leq 1/(k_{\textup{F,res}} a) \leq -0.5$, corresponding to a mean value of $0.24\mu$.% as stated in the main text. For the data in the unitary and the BEC regime, where we observe no spin currents within error bars, $\Delta b$ is comparable to $\mu$ itself.

\section{Extraction of the conductances}
$G_N$ is determined within linear response as in \cite{krinner_observation_2015}.
We 
%linearize around equilibrium the equations relating current to chemical potential bias on the one hand, $I_N = -\frac{1}{2} \frac{d}{dt} \Delta N =G_N\Delta\mu$; and particle number to chemical potential on the other hand, $\Delta N = \kappa \Delta \mu$ ($\kappa$ denoting the compressibility of a single reservoir at equilibrium, for a single spin component). The temporal evolution of $\Delta N$ is then governed by the differential equation
use the relation $\Delta N = \kappa \Delta \mu$, yielding
\begin{equation}\label{diffEqu}
\frac{\text{d}}{\text{d}t}\Delta N=-\frac{2G_N}{\kappa}\Delta N.
\end{equation}
We indeed observe an exponential decay of $\Delta N$ as a function of time (except for the deep superfluid regime, see SI Text). The characteristic time $\tau_N$ is related to $G_N$ through
$G_N = \kappa/2\tau_N$. 
To determine $G_N$, we evaluate $\kappa$ (see above), and determine $\tau_N$ by measuring $\Delta N$ at $t=0$ and after a transport time of $t_{\text{tr}}=2s$. From the solution of Eqn. (\ref{diffEqu}) we obtain
\begin{equation}
\frac{1}{\tau_N}=\frac{1}{t_{\text{tr}}}\ln\left(\frac{\Delta N}{N}(t=0)\right)-\frac{1}{t_{\text{tr}}}\ln\left(\frac{\Delta N}{N}(t=t_{\text{tr}})\right). \label{eqn:tauN}
\end{equation}
$G_\sigma$ is extracted slightly differently: a linear response relation similar to Eqn. (\ref{diffEqu}) cannot be established for $\Delta M$ because the magnetic susceptibility depends in a non-linear way on $b$. In particular, $\chi$ starts close to zero for low $b$ due to the superfluid gap. %(in a homogeneous system at zero temperature, $\chi$ is strictly zero below the critical polarisation). 
We define $G_\sigma $ as $G_\sigma = \frac{I_\sigma}{\Delta b}$.
We evaluate $\Delta b$ from the initial magnetization imbalance $\Delta M_0$ (see above) and determine $I_\sigma$ from
\begin{equation}
I_\sigma = \Delta M_0/(2\tau_\sigma),
\end{equation}
where $\tau_\sigma$ is the time constant of the observed exponential decay of the magnetization imbalance $\Delta M(t)$. $\tau_\sigma$ is determined by measuring $\frac{\Delta M}{N}$ at $t=0$ and after a transport time of $t_{\text{tr}}=4s$, and evaluating 
\begin{equation}
\frac{1}{\tau_\sigma}=\frac{1}{t_{\text{tr}}}\ln\left(\frac{\Delta M}{N}(t=0)\right)-\frac{1}{t_{\text{tr}}}\ln\left(\frac{\Delta M}{N}(t=t_{\text{tr}})\right). \label{eqn:tauSigma}
\end{equation}

\section{Error bars}
Error bars in Fig. \ref{fig3} and \ref{fig2} are statistical and indicate one standard deviation. Error bars in Fig. \ref{fig5} represent the uncorrelated combination of one standard deviation statistical and systematic uncertainties. The systematic uncertainty in the conductance amounts to 11\%. It represents the uncorrelated combination of the uncertainties in the compressibility, which are due to the calibration error in the total particle number, an uncertainty in the overall trapping potential and an uncertainty due to the use of the zero temperature equation of state. The statistical error in $T/T_c$ is due to the determination of $T$ and amounts to 10\%. The systematic uncertainty in $T/T_c$ is mainly due to the uncertainty in our estimate of $T_c$. It is caused by the overall uncertainty in the effective potential, which is due to the uncertainties in $\nu_z$, $\nu_x$, $V_g$, and their spatial dependencies. None of these uncertainties could explain the departure of conductance from $1/h$ observed in Fig.\ref{fig2} and \ref{fig5}.

\begin{acknowledgments}
We thank Shuta Nakajima for experimental assistance and G. Haack, J. Blatter, T. Giamarchi, J. von Delft, L.Glazman, N. Dupuis and W. Zwerger for discussions, and M. Landini, P.T\"{o}rm\"{a} and E. Demler for their careful reading of the manuscript and for discussions. We acknowledge financing from NCCR QSIT, the ERC project SQMS, the FP7 project SIQS, the Horizon2020 project QUIC, Swiss NSF under division II. JPB is supported by the Ambizione program of the Swiss NSF.
\end{acknowledgments}

\bibliographystyle{pnas}
%\bibliography{paper}

\pagebreak

%%%%%%%%%%%%%%%%%%%%%%%%%%%%%%%%%%%%%%%%%%%
%% SUPPLEMENTARY MATERIAL
%%%%%%%%%%%%%%%%%%%%%%%%%%%%%%%%%%%%%%%%%%%

\renewcommand{\figurename}{Figure}
\renewcommand{\thetable}{\arabic{table}} 
\setcounter{equation}{0}
\setcounter{figure}{0}
\renewcommand{\theequation}{S\arabic{equation}}
\renewcommand{\thefigure}{S\arabic{figure}}
\renewcommand{\bibnumfmt}[1]{[S#1]}
\renewcommand{\citenumfont}[1]{S#1}

\section*{Supplementary Information: Quantum simulation of spin and particle conductances in a quantum point contact}

\section{Temperature measurement}
\begin{figure*}
    \centering
    \includegraphics{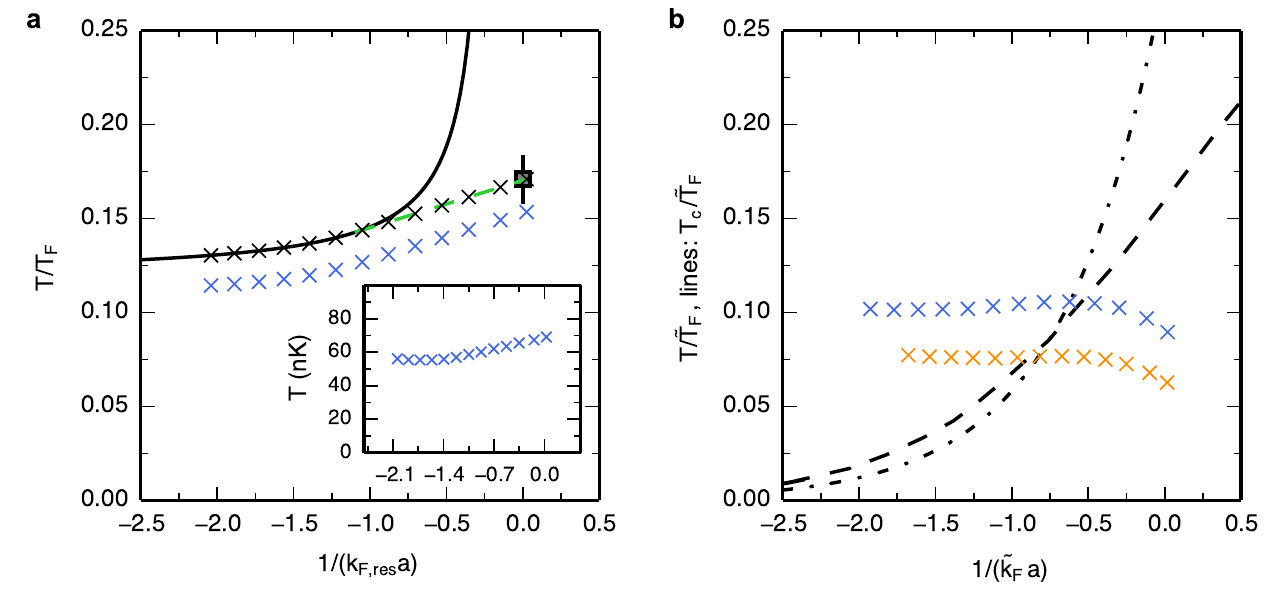}
    \caption{{\bf Temperature estimate from the BCS to the unitary regime.} \textbf{a,} Solid black line: Prediction of Eqn. (\ref{eqn:entropy}) based on the extracted entropy $S/N\textup{k}_{\textup{B}}=1.17(15)$ of the unitary gas, corresponding to $T/T_{\textup{F}}=0.17(1)$ indicated by the black square in the graph. Dashed green line: linear interpolation between the BCS prediction and the unitary regime (see text). Black crosses: $T/T_F$ of the sampled interaction strengths. Blue crosses: same as black crosses, but with an amount of off-resonant heating from the dipole trap subtracted, which occurs between the middle of the transport process and the imaging. \textbf{b,} Ratio of temperature to the local Fermi temperature $\tilde{T}_{\textup{F}}$ as a function of the local interaction strength $1/(\tilde{k}_{\textup{F}}a)$ (see text). Blue crosses correspond to the local $T/\tilde{T}_{\textup{F}}$ and $\tilde{k}_{\textup{F}}$ at the trap centre in the absence of the QPC, and are obtained from the blue crosses in subpanel a,. Orange crosses correspond to the local $T/\tilde{T}_{\textup{F}}$ and $\tilde{k}_{\textup{F}}$ at entrance and exit of the QPC for the parameters for which the centre of the conductance plateau in Fig. 3b of the Main Text is reached.
    The dashed black line is the ratio of critical temperature to Fermi temperature predicted by \cite{haussmann_thermodynamics_2007}, and the dash-dotted line corresponds to BCS theory with Gorkov and Melik-Barkhudarov corrections.}
    \label{fig:temperatures}
\end{figure*}

For the temperature measurement we ramp the magnetic field back to 689\,Gauss and use the virial theorem valid for a unitary gas~\cite{thomasvirial2005}
to determine the internal energy from the second moment of the density distribution. The temperature of the unitary gas is obtained from 
the internal energy via the known equation of state \cite{ku_revealing_2012, guajardo_higher-nodal_2013}. To trace back the temperature in the BCS regime, we suppose 
that the entropy remains constant during the magnetic field sweep, which is based on our observation that the temperature of the unitary gas is 
independent of the interaction strength set during transport. We thus set the entropy of the unitary gas, also extracted with the help of the equation of state, equal to the entropy of a gas in the BCS regime, which is given by \cite{carr_achieving_2004}
\begin{equation}
S_{\textup{BCS}} = \textup{k}_{\textup{B}}N\pi^2 T/T_{\textup{F}}\times\left(1+\frac{64 k_{\textup{F,res}}a}{35\pi^2}\right)\label{eqn:entropy}
\end{equation}
Temperatures extracted in this way are shown in Fig. \ref{fig:temperatures}a. The solid black line is obtained from Eqn. (\ref{eqn:entropy}) using the extracted entropy $S/N\textup{k}_{\textup{B}}=1.17(15)$ of the unitary gas. It corresponds to $T/T_{\textup{F}}=0.17(1)$ indicated by the black square in the graph. Since Eqn. (\ref{eqn:entropy}) contains interaction effects only to lowest order, we linearly interpolate the temperature for interaction strengths $-1.1 \leq 1/(k_{\textup{F,res}}a) \leq 0$ (green dashed line in Fig. \ref{fig:temperatures}a) between the prediction of Eqn. (\ref{eqn:entropy}) for $1/(k_{\textup{F,res}}a)=-1.1$ and the measured value at unitarity. Here, the upper limit of the validity range of Eqn. (\ref{eqn:entropy}), $-\infty \leq 1/(k_{\textup{F,res}}a) < -1.1$, has been chosen such that the equation predicts a value of $T/T_{\textup{F}}$ which is maximally by 20\% larger than its value for $1/(k_{\textup{F,res}}a)\rightarrow-\infty$.

Black crosses in Fig. \ref{fig:temperatures}a mark the interaction strengths that have been sampled in the experiment. The adiabatic sweep from the unitary regime to the BCS regime reduces $T/T_{\textup{F}}$ by $\sim25\%$ at the weakest value of the interaction strength in the BCS regime, which is $1/(k_{\textup{F,res}}a) = -2.1$. For $1/(k_{\textup{F,res}}a)=-1.9$ the temperature has been independently estimated from a degenerate Fermi gas fit to the 2D density profile which was acquired by keeping the magnetic field for imaging at the same value as for the transport process. It yielded $T/T_{\textup{F}}=0.13(2)$ in perfect agreement with the value deduced from the entropy of the unitary gas.

The temperatures stated so far correspond to the temperatures at the end of the experimental sequence where the absorption pictures are taken. Since in our experimental sequence some time elapses between the transport process and the imaging, we have to correct those temperatures for heating occurring in this time interval (typically $\sim 8\,$nK) in order to get the actual temperatures during the transport process. 
They are plotted as blue crosses in Fig. \ref{fig:temperatures}a. The absolute temperatures are shown in the inset.

Since predictions for the critical temperature in the regime between BCS and unitarity are usually made for uniform gases, we convert $T/T_{\textup{F}}$ into $T/\tilde{T}_{\textup{F}}$ and $1/(k_{\textup{F,res}}a)$ into $1/(\tilde{k}_{\textup{F}}a)$ (see Main Text). For their computation we need to know the density at the trap centre (in the absence of the QPC), which we estimate from the known zero temperature equation of state (see above). The resulting values are plotted as blue crosses in Fig.\ref{fig:temperatures}b as a function of $1/(\tilde{k}_{\textup{F}}a)$. Compared to Fig.\ref{fig:temperatures}a these temperatures are shifted towards lower values because $\tilde{T}_{\textup{F}}$ increases with density, which increases due to attractive interactions. 

We now turn to the local $T/\tilde{T}_{\textup{F}}$ and $\tilde{k}_{\textup{F}}$ at entrance and exit of the QPC. We obtain them by calculating the local density at those points using the known effective potential and the equation of state. 
Orange crosses show those values for the parameters for which the centre of the conductance plateau is reached in Fig. 3b and d, i.e. for $V_g=0.64\,\upmu$K and $\nu_x=23.2$\,kHz. Note that in the evaluation of the density, we account for the zero-point energy of the 2D confinement by subtracting $h \nu_z f_z(y)/2$ from the local chemical potential. While this is correct for the weakly attractive case, this correction is probably overestimated in the strongly interacting case \cite{levinsen_strongly_2015}. 
For the parameters for which the centre of the conductance plateau is reached in the data of Fig. 3a and c, i.e. for $V_g=0.42\,\upmu$K and $\nu_x=14.5$\,kHz, the effective potential at entrance and exit of the QPC turns out to be close to zero, and hence the corresponding $T/\tilde{T}_{\textup{F}}$ and $\tilde{k}_{\textup{F}}$ are represented by the blue crosses

The dashed black line in Fig.\ref{fig:temperatures}b is the ratio of critical temperature to Fermi temperature as predicted by \cite{haussmann_thermodynamics_2007}. The dash-dotted line is the prediction of BCS theory including the Gorkov and Melik-Barkhudarov corrections \cite{pethick2002bose}, $T_{\textup{c}}/\tilde{T}_{\textup{F}} = 0.28\, \textup{exp}[\pi/(2\tilde{k}_{\textup{F}}a)]$.

\section{Series spin resistance in the reservoirs}
Interactions between the two spin components in the reservoirs lead to a diffusive spin transport in the reservoirs. Here we estimate the effect of this spin diffusion on the measured spin conductance, by modelling the system as consisting of the QPC in series with the reservoirs, to which we assign a spin resistance $1/G_{\sigma \mathrm{R}}$. The measured conductance is then 
\begin{equation}
G_{\sigma} = \frac{1}{2/G_{\sigma \mathrm{R}} + 1/G_{\sigma \mathrm{QPC}} }.
\end{equation}

To estimate $G_{\sigma \mathrm{R}}$, we use the Drude model, stating that the conductivity is $\sigma = \frac{n \tau}{m}$, with the density $n$ and the scattering time $\tau=\frac{1}{n \sigma_s v_F} \left( \frac{T}{T_F} \right)^{-2}$. In the latter expression, $\sigma_s$ is the scattering cross section, $v_F$ is the Fermi velocity and the factor $\left( \frac{T}{T_F} \right)^{-2}$ accounts for Pauli suppression of scattering at low temperature. 

Considering the reservoir as a box with section $A$ and length $L$, the series resistance $1/G_{\sigma \mathrm{R}}$ can be written as $\frac{L}{\sigma A}$. To facilitate comparison with $1/G_{\sigma \mathrm{QPC}}\sim h$, we introduce a trap frequency $\omega$ corresponding to the level spacing in the QPC, and the corresponding harmonic length $l_{\mathrm{ho}} = \sqrt{\frac{\hbar}{m \omega}}$. After simple algebra, we obtain
\begin{equation}
\frac{1}{G_{\sigma \mathrm{R}}} =  h \frac{1}{\pi \sqrt{2}} \frac{\sigma_s L}{A l_{\mathrm{ho}}} \left( \frac{\mu_{\mathrm{res}} + V_g}{\hbar \omega} \right)^{-3/2} \cdot \left( \frac{k_B T}{\hbar \omega} \right)^2,
\end{equation}
where we have replaced the Fermi energy by $\mu_{\mathrm{res}} + V_g$, which is the sum of the reservoirs' chemical potential and the gate potential. Note that even though the expression formally contains $\omega$, it is actually independent of the QPC trap frequency. Taking reasonable parameters for the experiment, $L = \sqrt{A} \sim 30\,\mu$m, a scattering length of $-10000\,$a$_0$, a temperature $\frac{kT}{\hbar \omega}\sim 0.1$ and a gate potential such that $\frac{\mu_{\mathrm{res}} + V_g}{\hbar \omega} \sim 1$, we obtain $G_{\sigma \mathrm{R}} > \frac{100}{h}$.
We thus expect the effects of the spin resistance of the reservoirs to be negligible for a QPC in the single mode regime, and we disregard $1/G_{\sigma R}$ in the interpretation of the data.

For the unitary Fermi gas, the scaling is expected to be different since $\sigma_s$ is inversely proportional to the Fermi wavelength rather than a constant. In this regime, we can estimate the scattering time $\tau \sim \frac{h}{E_F} \left( \frac{T}{T_F} \right)^{-2}$ from dimensional analysis and accounting for the Pauli suppression of scattering. Taking the same estimates for the chosen geometry yields an even lower estimate for the series resistance.

\section{Model for the spin conductance in the superfluid phase}
In the superfluid phase with spin imbalanced populations, the mean-field Hamiltonian at each point $y$ along the 
transport direction (see Fig.~2c in the Main Text) reads~\cite{pethick2002bose}:
\begin{equation}
 \mathcal{H}_{mf} = \sum_k \xi_{k,s}(y)\gamma^\dagger_{k,s}\gamma_{k,s}\,,
\end{equation}
with $\gamma_{k,s}$ the Bogoliubov quasiparticle with momentum $k$ and spin $\sigma=\pm 1$, and a dispersion
relation $\xi_{k,\sigma}(y) = \sigma\cdot b \pm \sqrt{\left(\frac{\hbar^2k^2}{2m}-\mu\right)^2+\Delta(y)^2}$. Since 
the paired particles carry no spin, the only contribution to spin transport originates from the 
excitations $\gamma_{k,\sigma}^\dagger$ on top of the superfluid background. Then, spin transport can be viewed as the scattering of
the Bogoliubov particles generated in the reservoirs located at $y=\pm\infty$ (with an energy just 
above the gap $\Delta_\text{res}$ in the reservoirs) through a potential barrier representing the 
space-dependent spin gap $\Delta(y)$. This approach is valid for the spin current because it is entirely 
carried by the normal fraction of the gas, in contrast to the particle current which has a genuine superfluid 
contribution. Assuming local thermodynamic equilibrium and applying the WKB approximation, 
the probability for a Bogoliubov quasiparticle to be scattered from one reservoir to the other is given by:
\begin{equation}
 \label{eq:probability_WKB}
 \mathcal{T}(\varepsilon,V_g) = \left| \exp{\left[-\frac{\sqrt{2m}}{\hbar}\int_{-\infty}^{+\infty} dy \sqrt{\Delta(y,V_g)-\varepsilon}\right]}\right|^2\,.
\end{equation}

The spin current is obtained as the sum of the contributions from both branches of the spectrum of Bogoliubov quasiparticles~:
\begin{align}
\label{eq:spincurrent}
 &I_\sigma = \frac{1}{h} \int_{\Delta_\text{res}}^{+\infty} d\varepsilon \Phi(\varepsilon, V_g)\frac{ (f_{\uparrow,L}-f_{\downarrow,L})(\varepsilon) - 
 (f_{\uparrow,R}-f_{\downarrow,R})(\varepsilon)}{2} \\
 & + \frac{1}{h}\int_{-\mu}^{-\Delta_\text{res}} d\varepsilon \Phi(\varepsilon, V_g)\frac{ (f_{\uparrow,L}-f_{\downarrow,L})(\varepsilon) - 
 (f_{\uparrow,R}-f_{\downarrow,R})(\varepsilon)}{2} \,,
\end{align}
with the transport function $\Phi(\varepsilon, V_g) = \sum_n \mathcal{T}(\varepsilon, V_g)\vartheta(\varepsilon - (E_n-V_g-\mu))$,
which in the non-interacting limit counts the number of transport channels 
available for a particle having an energy $\varepsilon$. In~\eqref{eq:spincurrent}, 
$f_{\sigma,L(R)} = \frac{1}{1+\exp{\left[(\varepsilon-\sigma\cdot b_{L(R)})/k_BT\right]}}$ is the 
distribution of Bogoliubov particles with spin $\sigma$ in the left (right) reservoir. It
immediately shows that the population of excitations carrying spin is driven by the spin potentials
$b_{L(R)} = \frac{\mu_{\uparrow,L(R)}-\mu_{\downarrow,L(R)}}{2}$.\\
Within linear response, and in a configuration such that $b_L = -b_R \equiv b$ (symmetric spin bias), one can rewrite the spin current as:
\begin{align}
\label{eq:spincurrent_LR}
I_\sigma =& \frac{2b}{h} \Biggl[\int_{\Delta_\text{res}}^{+\infty} d\varepsilon \Phi(\varepsilon) \left(-\frac{\partial f_{\uparrow,L}}{\partial \varepsilon}\right)_{b=0} \\
& +\int_{-\mu}^{-\Delta_\text{res}} d\varepsilon \Phi(\varepsilon) \left(-\frac{\partial f_{\uparrow,L}}{\partial \varepsilon}\right)_{b=0}  \Biggr]\,.
\end{align}

Inserting in~\eqref{eq:spincurrent_LR} the expression of the transport function gives the following 
expression for the spin conductance:
\begin{align}
 \label{eq:spin-conductance_theory}
 G_\sigma =& \frac{1}{h} \sum_n \Biggl[ \int_{\Delta_\text{res}}^{+\infty} d\varepsilon \frac{\mathcal{T}(\varepsilon, V_g)\vartheta(\varepsilon-(E_n-V_g-\mu))}{4 k_B T \cosh^2{\left[\frac{\varepsilon}{2k_BT}\right]}} \\
 & +\int_{-\mu}^{-\Delta_\text{res}} d\varepsilon \frac{\mathcal{T}(\varepsilon, V_g)\vartheta(\varepsilon-(E_n-V_g-\mu))}{4 k_B T \cosh^2{\left[\frac{\varepsilon}{2k_BT}\right]}}\Biggr],
\end{align}
which in the non-interacting regime reduces to the usual expression for the conductance of a QPC. 
The $\vartheta$-function in Eqn.~\eqref{eq:spin-conductance_theory} tends to increase the conductance as $V_g$ is raised, whereas the term $\mathcal{T}(\varepsilon, V_g)$ is responsible for an exponential suppression of $G_\sigma$ with increasing $V_g$ because the spin gap is proportional to the local Fermi 
energy. For quantitative comparison, we take the following
expression of the superfluid gap~\cite{pethick2002bose}~:
\begin{equation}
\label{eq:gap_Gorkov}
 \Delta(y) = 0.493 E_F(y) \cdot \exp{\left[\frac{\pi}{2k_F(y)a}\right]}\,.
\end{equation}

\section{Spin drag}

\begin{figure}
\centering
    \includegraphics[width=89mm]{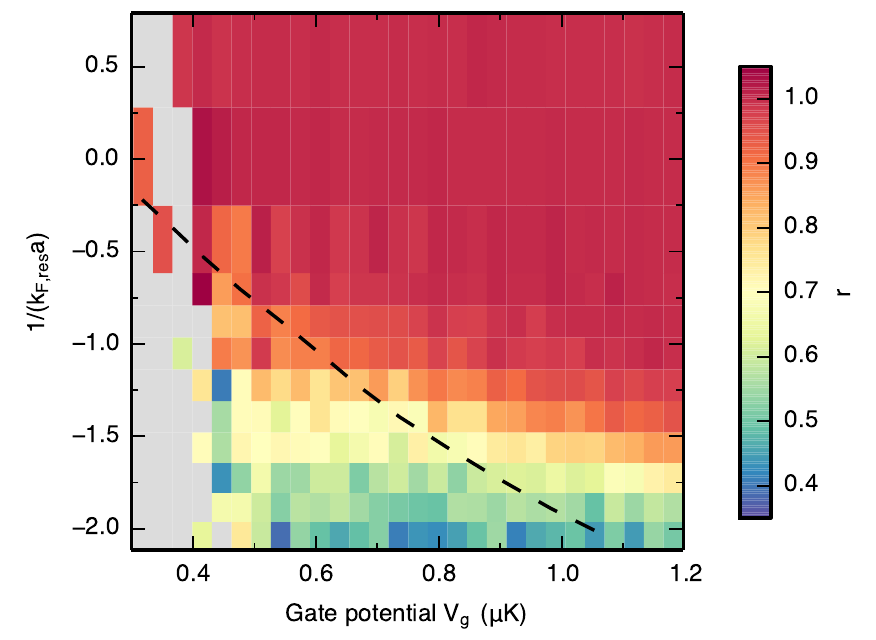}
    \caption{{\bf Comparison between particle and spin conductances.}
    Ratio $r = \frac{G_N-G_\sigma}{G_N+G_\sigma}$ as a function of gate potential $V_g$ and interaction strength 
    in the reservoirs. Points where $G_N < 0.05/h$, corresponding to a pinched-off QPC, are not shown. The dashed line indicates the expected superfluid transition.}
    \label{fig4}
\end{figure}

For a mixture of two interacting spin components, the currents for individual spin components are given within linear response by
\begin{equation}
\begin{cases}
 I_\uparrow  =  G_\uparrow \Delta\mu_\uparrow + \Gamma \Delta\mu_\downarrow \\
 I_\downarrow = \Gamma \Delta\mu_\uparrow + G_\downarrow \Delta\mu_\downarrow
 \end{cases}
 \label{eq:spinDrag}
\end{equation}
where $I_\sigma$ is the current of spin $\sigma$ atoms and $\Gamma$ is a coefficient describing spin drag~\cite{Spindrag_Flensberg}. In particular, the previous relations yield
\begin{equation}
 I_\uparrow = \left[G_\uparrow-\frac{\Gamma^2}{G_\downarrow}\right]\Delta\mu_\uparrow+\frac{\Gamma}{G_\downarrow}I_\downarrow
\end{equation}
The quantity $\Gamma/G_\downarrow$ can indeed be interpreted as the fraction of particles of spin $\uparrow$ 'dragged' by the flow of particles of spin $\downarrow$.\\

In the absence of a global spin polarisation we have $G_\uparrow=G_\downarrow\equiv G$, and Eqn.~(\ref{eq:spinDrag}) is diagonalised in the particle-spin basis. We obtain 
\begin{equation}
\begin{cases}
I_N = \frac{1}{2}\left( I_\uparrow + I_\downarrow \right) = G_N \Delta \mu \\
I_\sigma = \frac{1}{2}\left( I_\uparrow - I_\downarrow \right) = G_\sigma \Delta b,
\end{cases}
\end{equation}
with
\begin{equation}
\begin{cases}
G_N = G + \Gamma \\
G_\sigma = G - \Gamma.
\end{cases}
\end{equation}

We characterise the spin drag by the ratio $r = \frac{G_N-G_\sigma}{G_N+G_\sigma} = \frac{\Gamma}{G}$.
In the non-interacting regime, $\Gamma$ is zero and so is $r$, while for a maximally correlated flow,
$G_\sigma$ is zero and $r = 1$. 
Note that for magnetic insulators such as produced with repulsive Fermions in optical lattices, 
$r$ would be close to $-1$.

We use the data of Fig. 2b and Fig. 3d to extract $r$. The result is plotted as a function of interaction 
strength and gate potential in Fig. \ref{fig4}. In the regime of high gate potential and strong interactions, $r$ is equal to one: the spin channel is 
entirely closed by pairing and particle transport proceeds only via singlet pairs. For weaker interactions or lower gate potentials, $r$ decreases to about $0.3$ but remains significantly larger than zero. 
The non-zero value of $r$ illustrates the fact that contrary to the particle transport, spin transport turns diffusive and is reduced by interparticle scattering, as collisions transfer momentum between the two spin components. In one-dimensional systems, this process is responsible for spin-charge separation.

\section{Mean field reduction of the plateau width}
\begin{figure}
    \includegraphics[width=89mm]{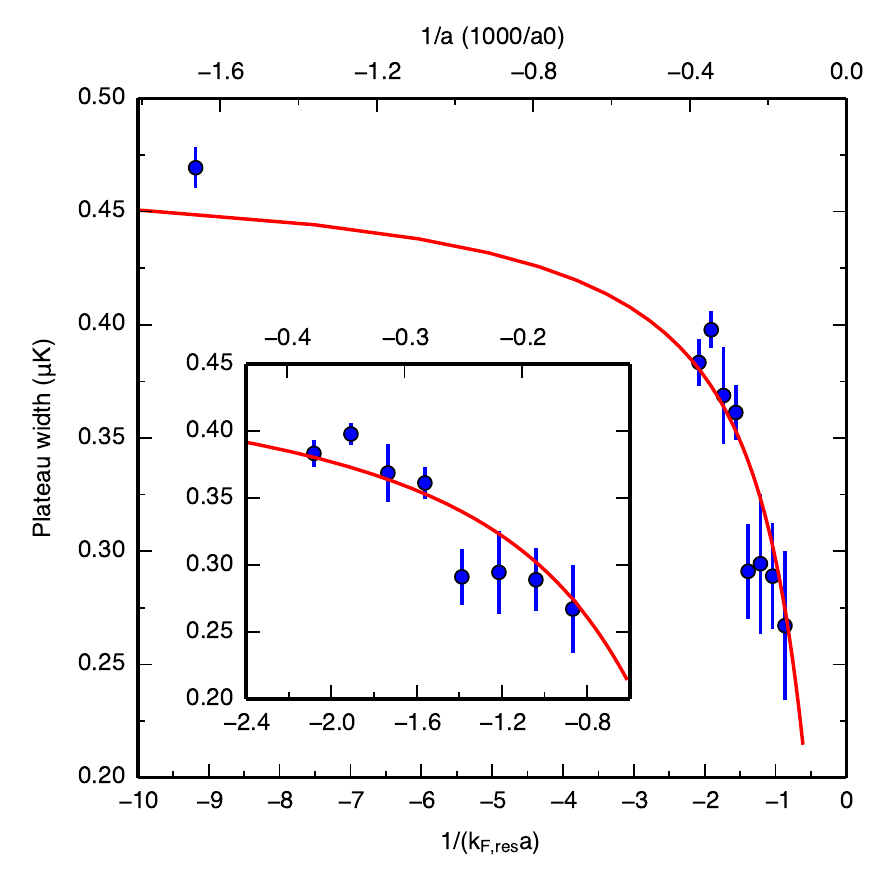}
    \centering
    \caption{{\bf Plateau width as a function of interaction strength.} The values are extracted from sigmoid fits to the data of Fig. 3b and d. An additional reference point at $1/(k_{\textup{F,res}} a)=-9$ is obtained from a measurement with a weakly interacting Fermi gas at a magnetic field of $491\,$G. Inset: zoom on the interaction strength studied in this work.}
    \label{fig:plateauWidth}
\end{figure}

For the data set of Fig. 3b and d we extract the plateau width of those conductance curves, for which we observe clear plateaux or a 
remaining bending of the curve, i.e. for interaction strengths $1/(k_{\textup{F,res}} a) \leq -0.9$. We do so by fitting the sum of two sigmoid functions having equal 
steepness and amplitude.
Fig. \ref{fig:plateauWidth}a shows the fitted plateau widths as a function of $1/(k_{\textup{F,res}} a)$ and $1/a$. The data point to the very 
left is extracted from a reference measurement using a weakly interacting Fermi gas prepared on the other side of the Feshbach resonance at a magnetic field of $491\,$G, where $a=-601\,a_0$. 
The data is in good agreement with a mean-field model of the QPC (solid red line in Fig. \ref{fig:plateauWidth}), 
which includes intra- and inter-mode interactions on a mean field level to determine the occupation of the transverse modes in a self-consistent way. The conductance is then calculated using the Landauer formula and the resulting theory curves are fitted with the same sigmoid fit functions as the data. Attraction between particles of the ground state and the first excited mode leads to an occupation of the latter at an already lower gate potential, and thus to a decreased plateau width. Intra-mode interactions on the other hand increase the steepness of the conductance rise to the plateau values. This effect however is rather small and not resolved in our measurements.

\section{Non-linear response of particle currents in the superfluid regime}

\begin{figure*}
    \centering
    \includegraphics[width=88.0mm]{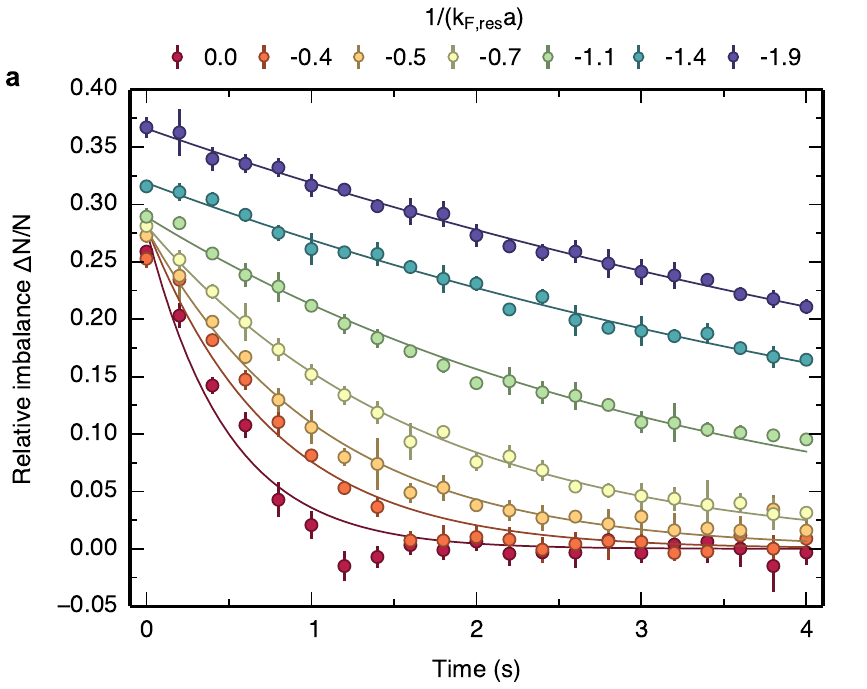}
    \includegraphics[width=64.0mm]{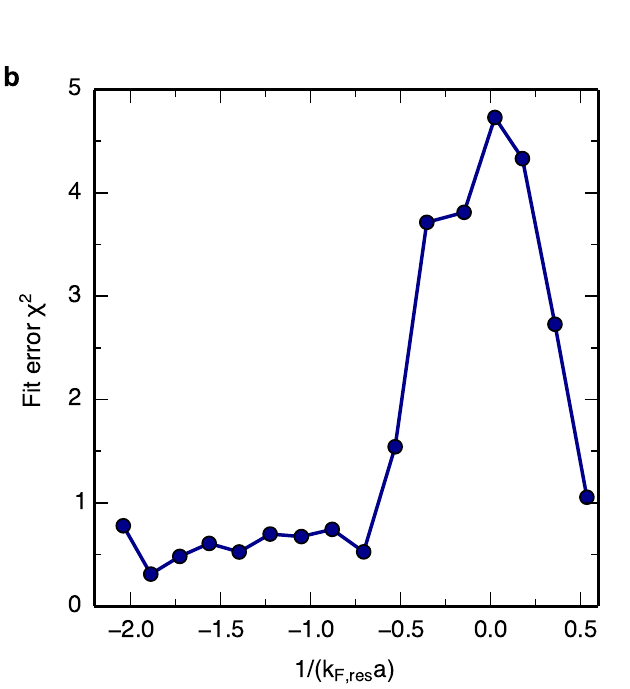}
    \caption{{\bf Time evolution of the particle number imbalance.} \textbf{a,} Relative imbalance between the two reservoirs as a function of time for various values of the interaction strength $1/(k_{\textup{F,res}}a)$ in the reservoirs. The horizontal confinement and the gate potential were set to $\nu_x=23.2\,$kHz and $V_g=0.64\,\upmu$K respectively, corresponding to the centre of the conductance plateaux in Fig. 3b and d. Solid lines are exponential fits. \textbf{b,} Fit error $\chi^2$ as a function of interaction strength. The sharp rise at $1/(k_{\textup{F,res}}a)\sim-0.7$ indicates the onset of non-linear current-bias characteristics in the strongly interacting regime.}
    \label{fig:decay}
\end{figure*}

For a fixed gate potential and horizontal confinement, $V_g = 0.67\,\mu$K and $\nu_x = 23.2\,$kHz, corresponding to the centre of the conductance plateaux of Fig. 3b and d, we measure the 
decay of the initial particle imbalance as a function of time. A linear relation between the current 
and the chemical potential bias implies an exponential decay, like for the discharge of a capacitor 
in a RC circuit. Deviations from exponential behaviour signal a breakdown of linear response and appear in the deep superfluid regime \cite{Husmann:2015aa}.

Fig. \ref{fig:decay} shows a set of experimental decay curves, together with an exponential 
least-square fit. For a wide range of interaction strengths, $1/(k_{\textup{F,res}}a)<-0.5$, the data is adequately 
fitted by an exponential. For stronger interactions, systematic deviations from an exponential appear. 

We assess the quality of the exponential fit using the reduced $\chi^2$ obtained from the least square 
method. Fig. \ref{fig:decay}b presents the evolution of $\chi^2$ as a function of interaction strength. 
It shows a sharp increase for $1/(k_{\textup{F,res}}a)>-0.5$,
which coincides with the transition to the deep purple region in Fig. 3c.

In this regard, the meaning of $G_N$ as calculated through Eqn. (8) and (9) in Materials and Methods is less straightforward. Eqn. [9] can be rewritten as:
\begin{equation}
\frac{1}{\tau_N} = -\frac{1}{t_r} \int_0^{t_r} \frac{\frac{d}{dt} \Delta N(t)}{\Delta N(t)} dt
\end{equation}
Using the thermodynamic relation $\Delta N = \kappa \Delta \mu$ and the definition of the current in Eqn. (\ref{eq:linearresp}), Eqn. (\ref{eqn:estimatedGN}) reads
\begin{equation}
G_N = \frac{1}{t_r} \int_0^{t_r} \frac{I_N(t)}{\Delta \mu(t)} dt.
\end{equation}
The estimated conductance appears as a time average of the particle current-to-bias ratio. Assuming that $I_N$ is a concave function of $\Delta \mu$ (as observed in \cite{Husmann:2015aa} at unitarity), $G_N$ is therefore a lower bound for the linear conductance defined as $\left(\frac{d I_N}{d\Delta\mu}\right)_{\Delta\mu = 0}$.

\end{document}